\documentclass{elsart}
\usepackage{epsfig}
\usepackage{amssymb}
\usepackage{graphicx}
\usepackage{bm}
\def\simge{\mathrel{%
   \rlap{\raise 0.511ex \hbox{$>$}}{\lower 0.511ex \hbox{$\sim$}}}}
\def\simle{\mathrel{
   \rlap{\raise 0.511ex \hbox{$<$}}{\lower 0.511ex \hbox{$\sim$}}}}

\def\lapproxeq{\lower .7ex\hbox{$\;\stackrel{\textstyle
<}{\sim}\;$}}
\def\gapproxeq{\lower .7ex\hbox{$\;\stackrel{\textstyle
>}{\sim}\;$}}
\def \pom {{\hspace{ -0.1em}I\hspace{-0.2em}P}}
\def \reg {{\hspace{ -0.1em}I\hspace{-0.2em}R}}
\newcommand{\be}{\begin{equation}}
\newcommand{\ee}{\end{equation}}
\newcommand{\beqn}{\begin{eqnarray}}
\newcommand{\eeqn}{\end{eqnarray}}
\begin{document}

\begin{frontmatter}

\title{Charm production in diffractive deep inelastic scattering}

\author{V.P. Gon\c{c}alves$^{1}$ and M.V.T. Machado$^{1,2,3}$}
\address{ $^{1}$ Instituto de F\'{\i}sica e Matem\'atica,  Universidade
Federal de Pelotas\\
Caixa Postal 354, CEP 96010-090, Pelotas, RS, Brazil\\
$^{2}$ High Energy Physics Phenomenology Group, GFPAE IF-UFRGS \\
Caixa Postal 15051, CEP 91501-970, Porto Alegre, RS, Brazil\\
$^{3}$ Theory Division, CERN, CH-1211 Geneva 23, Switzerland}

\date{\today}

\begin{abstract}
The diffractive open charm production is computed in perturbative
QCD formalism and in the Regge approach. The results are compared
with recent data on charm diffractive structure function measured
at DESY-HERA. Our results demonstrate that this observable can be
useful to discriminate the QCD dynamics.

\end{abstract}
\end{frontmatter}

\section{Introduction}
The study of electroproduction at small $x$ has lead to the
improvement of our understanding of QCD dynamics at the interface
of perturbative and nonperturbative physics. In particular, the
discovery of diffractive events in this process at HERA has
triggered a large amount of experimental and theoretical work and
greatly increased our knowledge of the physics of diffraction (For
recent reviews see Refs. \cite{wusmartin,review1,PREDAZZI}).
 Diffractive
processes in deep-inelastic scattering (DIS) are of particular
interest, because the hard photon in the initial state gives rise
to  the hope that, at least in part, the scattering amplitude can
be calculated in pQCD. Moreover, DIS exhibits the nice feature of
having a colorless particle, the virtual photon, in the initial
state. The main theoretical  interest in diffraction is centered
around the interplay between the soft and hard physics. Hard
physics is associated with the well established parton picture and
perturbative QCD, and is applicable to processes for which a large
scale is present. Soft dynamics on the other hand, linked for
example with the total cross section of hadron scattering, is
described by nonperturbative aspects of QCD. The ability to
separate clearly the regimes dominated by soft and hard processes
is essential in exploring QCD at both  quantitative and
qualitative level.

Recently, we have proposed the analyzes of the slope of diffractive
structure function as a potential observable to disentangle the
leading dynamics at $ep$ diffractive processes \cite{plb,npa}. The
predictions for the behavior of this quantity are strongly
dependent of the QCD dynamics dominant in the kinematical region
(For a recent discussion see Ref. \cite{munier}). Similarly, the
study of the diffractive final state can lead to further progress
in the direction of obtaining a coherent picture of the diffraction.
In particular, charm production looks promising in this respect,
as predictions for this process  widely differ among several models  \cite{genovese,lotter,levin,diehl,haakman}. Recently, the
ZEUS collaboration has presented its results for the measurement
of the open-charm contribution to the diffractive proton structure
function \cite{zeus}. Consequently, a more detailed analyzes of
the models and a comparison between their predictions and the
experimental data is on time. Here the diffractive open charm
production is computed in perturbative QCD formalism and in the
Regge approach.  As these models are based on very distinct
assumptions, it allows shed light into the leading dynamics at
$ep$ diffractive processes.

This paper is organized as follows. In the next section, we
summarize the main formulas for computation of the open charm
diffractive structure function. One presents it in the transverse
momentum representation in the perturbative QCD approach.
Moreover, the diffractive production of open charm is calculated
in a Regge inspired approach, where charm is produced by boson
gluon fusion and which directly depends on the gluon distribution
of the  Pomeron. In the last section, we compare both approaches
with current  experimental measurements from DESY-HERA collider
and present our discussions and conclusion.

\section{Diffractive production of open charm in deep inelastic scattering }

Before introducing the main expressions needed to our calculation,
let's introduce the kinematical definitions in diffractive DIS
(DDIS) $\gamma^* \,(q) + p \,(P)\rightarrow X \,(M_{X})\,p\,(P')$
with $X$ being the diffractive final state.  The kinematics is
defined as follows,
\begin{eqnarray}
x=\frac {-q^2} {2P.q} \,,\hspace{0.7cm}   x_{\pom}=\frac {q.(P-P')} {q.P} \,, \hspace{0.7cm} \beta=\frac {-q^2} {2q.(P-P')}\approx \frac{Q^2}{Q^2+M_{X}^2}\,,
\end{eqnarray}
where  $q$, $P$ and $P'$ are the four-momenta of the virtual
boson, the incident proton and the remnant colorless final state,
respectively. The invariant mass of the diffractive final state is
labeled $M_X$. The variable $x$ is the momentum fraction of the
proton carried by the partons (quarks or gluons), the Bjorken
variable, and by definition  $x= \beta\, x_{\pom}$. As usual, 
$Q^2=-q^2$ is the photon virtuality.

At high  energies, $x_{\pom}$ may be interpreted as the fraction
of the proton four-momentum carried by the diffractive exchange,
the colorless Pomeron. The $\beta$ variable is the fraction of the
four-momentum of the diffractive exchange carried by the parton
interacting with the virtual boson. The diffractive structure
function is defined  in analogy with the decomposition of the
unpolarized total $ep$ cross section  as,
\begin{eqnarray}
\frac{d^3\sigma_{ep\rightarrow epX}}{dx_{\pom}\,d\beta \,dQ^2}=\frac{4\pi\alpha^2}{xQ^4}\left\{1-y+\frac{y^2}{2}\right\}F_2^{D(3)}(x_{\pom}, \beta, Q^2) \,.
\end{eqnarray}
Since the first observation of diffractive DIS at HERA, several
attempts have been made to compare the data with the Regge and
QCD-based models \cite{CKMT2,bartels,GBW} (See also
\cite{pescmun,royon1,royon2}). In general, these models provide a
reasonable description of the present data on the diffractive
structure function $F_2^{D(3)}$, although based on quite distinct
frameworks. Furthermore, the QCD factorization theorem has been
proven to be valid for $F_2^{D(3)}$ \cite{collins}, with the
immediate consequence that the DGLAP evolution equations
\cite{DGLAP} should describe the scaling violations observed in
this observable. However, there are no constraints on the gluon
momentum distribution since the momentum sum rule does not
formally apply.

Here, we study in detail  the predictions for the charm component
of the diffractive structure function,
$F_2^{D(3)\,\mathrm{charm}}$, which is directly sensitive to the
gluonic content of the pomeron, considering two distinct
approaches: i) a Regge inspired model \cite{CKMT1,CKMT2}, where
the diffractive production is dominated by  a nonperturbative
Pomeron, and the diffractive structure function is obtained using
the Ingelman-Schlein ansatz \cite{ingelman}. ii) a pQCD approach
\cite{bartels,GBW} where the diffractive process is modeled as the
scattering of the photon Fock states with the proton through a
gluon ladder exchange (in the proton rest frame). Below we present
a brief review of the main assumptions of these models.

In the   perturbative QCD framework, there are  successful analysis
describing the diffractive structure function \cite{bartels,GBW}.
The underlying physical picture is that, in the proton rest frame,
diffractive DIS is described by the interaction of the photon Fock
states ($q\bar{q}$ and $q\bar{q}g$ configurations) with the proton
through a Pomeron exchange, modeled as a two hard gluon exchange.
The corresponding structure function contains the contribution of
$q\bar{q}$ production to both the longitudinal and the transverse
polarization of the incoming photon and of the production of
$q\bar{q}g$ final states from transverse photons. The basic
elements of this approach are the photon light-cone wave function
and the nonintegrated gluon distribution (or dipole cross section
in the dipole formalism). For elementary quark-antiquark final
state, the wave functions depend on the helicities of the photon
and of the (anti)quark. For the $q\bar{q}g$ system one considers a
gluon dipole, where the pair forms an effective gluon state
associated in color to the emitted gluon and only the transverse
photon polarization is important. The interaction with the proton
target is modeled by two gluon exchange, where they couple in all
possible combinations to the dipole. Then the diffractive
structure function can be written as \cite{bartels,GBW}
 \begin{eqnarray}
F_2^D(x_{\pom},\beta,Q^2) \sim
\beta\,\int\,d\alpha\,\int\,\frac{k_t^2\,d^2k_t}{(1-\beta)^2} \,\,\left | \int \,\frac{d^2l_t}{l_t^2}\,D\Psi(\alpha,k_t) \, {\mathcal F}(x_{\pom}, l_t^2)\right
|^2\;, \end{eqnarray}
 where $D\Psi$ is a combination of the concerned wave
functions,  $l_t$ is the transverse momentum of the  exchanged
gluons. The function ${\mathcal F}(x_{\pom}, l_t^2)$ defines the
Pomeron amplitude (nonintegrated gluon distribution) and contains
all the details concerning the coupling of the $t$-channel gluons
to the proton. Integrating it over $l_t^2$ one obtains the
conventional collinear gluon distribution.

Concerning  diffractive open charm production, the  exclusive
$c\bar{c}$-pair arises from the dissociation of longitudinally and
transversely polarized photons, as well as the production of the
$c\bar{c}g$-state. The diffractive structure functions for
$\gamma^* p \rightarrow c\bar{c}\,p$ are given by
\cite{genovese,lotter,wusmartin}, \beqn
 x_{\pom}F_{T,c\bar{c}}^D(x_{\pom},\beta, Q^2) &  = & \frac{e_c^2}{48 B_D}\;       \frac{\beta}{(1-\beta)^2}\int \frac{dk_t^2}{k_t^2} \frac{k_t^2+m_c^2}{\sqrt{1-4\beta
k^2/Q^2}}\nonumber\\
 & \times & \Theta\left (k^2-\frac{Q^2}{4\beta}\right) \,
\left\{\left[1-\frac{2\beta k^2}{Q^2}\right]|I_T|^2 + \frac{4
k_t^2 m_c^2}{k^4}|I_L|^2\right\}, \label{FT} \\
 x_{\pom}F_{L,c\bar{c}}^D(x_{\pom}, \beta, Q^2) &  = &  \frac{e_c^2}{3 B_D Q^2} \int \frac{dk_t^2}{1-\beta} \!\frac{k^2\,\beta^3}{\sqrt{1-4\beta k^2/Q^2}}  \Theta\left(k^2-
\frac{Q^2}{4\beta}\right)\!|I_L|^2 \label{FL}
\eeqn
where the
upper  limit in the integration on the quark loop is constrained
by the $\Theta$-function. The parameter $B_D$ is the diffractive
slope, which arises by assuming a simple exponential form for the
$|t|$ dependence to the process (one uses $B_D=6$ GeV$^{-2}$ in
the following).  The integrals $I_{T,\,L}$ on gluon transverse
momentum are defined as, \beqn
 I_T  & = &   \int\frac{d\ell_t^2}{\ell_t^2} \alpha_s\,(\mu_c^2)\,{\mathcal F}(x_{\pom},\ell_t^2)
\left[1-2\beta-2\frac{m_c^2}{k^2}+ \frac{\ell_t^2 - (1-2\beta)\;k^2+2\;m_c^2}{\sqrt{(\ell_t^2
+ k^2)^2 -4 \ell_t^2 \;k_t^2}} \right] \label{IT} \nonumber\\
 I_L  & = &   \int\frac{d\ell_t^2}{\ell_t^2
} \alpha_s\, (\mu_c^2)\, {\mathcal F}(x_{\pom},\ell_t^2)
\left[1\;-\;\frac{k^2} {\sqrt{(\ell_t^2+k^2)^2 -4
\ell_t^2\;k_t^2}} \right]\;\;\,, \label{IL} \nonumber \eeqn where
the two-body kinematical relation has  a mass term (in relation to
the light quarks dipoles) and reads as
$k^2=\frac{k_t^2+m_c^2}{1-\beta}$. The allowed range on $\beta$ is
different from the light dipole case as the diffractive mass $M_X$
has a lower limit defined by $M_X^2\geq 4\,m_c^2$. For the energy
scale entering in the strong coupling we will use the prescription
$\mu_c^2=4m_c^2$.

In order to compute the contribution of the  $c\bar{c}g$
component,  one makes use of the  diffractive factorization
property \cite{wusmartin}. The diffractive gluon distribution
$g^D(\beta)$ will be convoluted  with the corresponding
charm-coefficient function $C_g(\zeta,r)$, \beqn F_{c\bar{c}g}^D
(x_{\pom}, \beta, Q^2) \label{fccg} &=& 2 \;\beta\; e_c^2\;
\frac{\alpha_s(\mu_c^2)}{4 \pi}\; \int_{a \beta}^1
\frac{dz}{z}\;C_g\left(\frac{\beta}{z},\frac{m_c^2}{Q^2}\right) \;
g^D(z) \label{f2dqqbarg} \eeqn where the lower limit in the $z$
integration is weighted by  $a=1+4m_c^2/Q^2$ and the coefficient
function is given by, \beqn
C_g\,(\zeta,r)& = &\left[\zeta^2+(1-\zeta )^2+4\zeta (1-3\zeta)r-8\zeta^2 r^2\right]\;\ln\frac{1+\varepsilon}{1-\varepsilon}\nonumber\\
& + & \varepsilon\left[-1+8\zeta (1-\zeta)-4\zeta
(1-\zeta)r\right]\,, \eeqn with $\varepsilon$,  the centre-of-mass
velocity of the charm quark or antiquark,  given by
$\varepsilon=\sqrt{1-(4r\zeta/1-\zeta)}$.

For the diffractive gluon distribution, we use the momentum
representation, which reads as \cite{wusmartin}
 \beqn \label{GD}
 g^D\,(\beta,x_{\pom}) & = &  \frac{9}{64 x_{\pom} B_D}\;\frac{1}{\beta\,(1-\beta)}\;\int dk_t^2\;\left\{\int\frac{d\ell_t^2}{\ell_t^2}\;\alpha_s (\mu_c^2)\,
{\mathcal F}(x_{\pom},\ell_t^2) \;\; \right. \nonumber \\
& \times &  \left.\left[\beta^2+(1-\beta)^2+\frac{\ell_t^2}{k^2}-
\frac{[(1-2\beta)k^2-\ell_t^2]^2+2\beta(1-\beta)k^4}
{k^2\sqrt{(\ell_t^2+k^2)^2-4(1-\beta)\;\ell_t^2\;k^2}}
\right]\right\}^2. 
\eeqn
The upper limit of the $k_t$-integration is fixed by the condition $M_X^2 = Q^2\frac{(1-\beta)}{\beta} > (k_t + \sqrt{k_t^2 + 4m_c^2})^2$, which implies $k_t^2  \lapproxeq \frac{Q^2}{4}\frac{(1-\beta)}{\beta}$.

The diffractive gluon distribution obtained above depends directly on
the unintegrated gluon function. Concerning the behavior on $\beta$,
an expansion in powers of $\ell_t^2/k^2$ \cite{wusmartin} produces $g^D\sim
\frac{1}{\beta}\,(1-\beta)^3(1+2\beta)^2\int dk_t^2\, [x_{\pom}g(x_{\pom},
k^2)]^2/k_t^4$, where $g(x_{\pom}, Q^2)$ is the collinear gluon
distribution. Therefore, the diffractive gluon distribution has a
singular behavior at $\beta \to 0$ and vanishes at $\beta \to 1$.

In order to perform further numerical analysis,  we will use the
unintegrated gluon function giving by the saturation model, which
has a simple analytical form \cite{GBW}. It reads as, \beqn
\alpha_s\,{\mathcal F}(x_{\pom},\ell_t^2) =
\frac{3\,\sigma_0}{4\pi^2}\,\left(\frac{\ell_t^2}{Q_{\mathrm{sat}}^2(x_{\pom})}
\right)\,\exp \left(-\frac{\ell_t^2}{Q_{\mathrm{sat}}^2(x_{\pom})}
\right)\,, \label{fbgw} \eeqn where
$Q_{\mathrm{sat}}^2(x)=\left(\frac{x}{x_0} \right)^{-\lambda}$,
$Q_{\mathrm{sat}}$ is the saturation scale and one has used the
parameters for the 4-flavor fit. Accordingly, for the computation
of the $q\bar{q}g$ contribution, Eq. (\ref{fbgw}) has been
properly rescaled concerning the color charge \cite{BJK}.

\begin{figure}[t]
\centerline{\epsfig{file=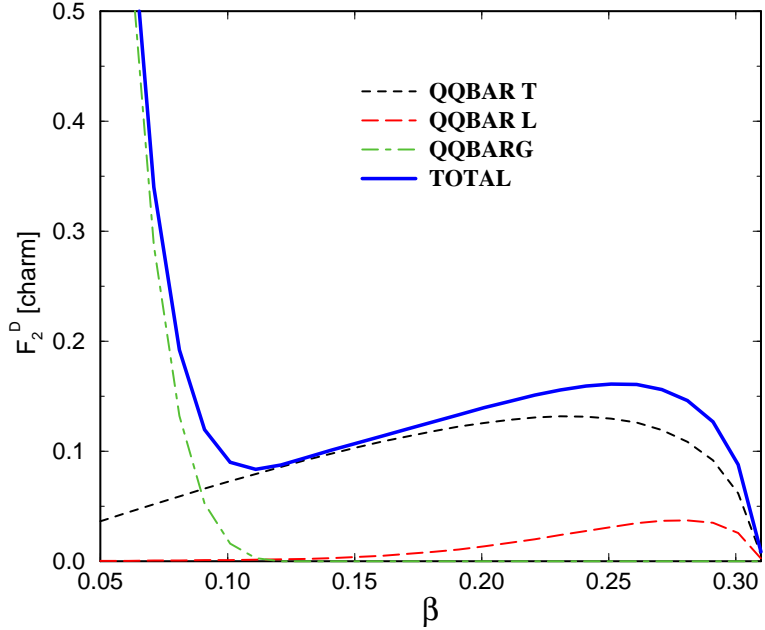,width=100mm}} \caption{The
open charm  diffractive structure function and its three
components plotted versus $\beta$ for a fixed $x_{\pom} = 0.004$
and $Q^2=$ 4 GeV$^2$. } \label{fig:1}
\end{figure}

In Fig. \ref{fig:1} we present how the three contributions,
$c\overline{c}g$, $c\overline{c}$ from transversely and
longitudinally polarized photons, contribute for the
$\beta$-spectrum of $F_2^{D(3)\,\mathrm{charm}}$. At small-$\beta$
we have that the $c\overline{c}g$ component dominates, which
implies that the fraction of charm in this regime is predicted to
be the same as expected in inclusive charm production ($\approx 25
\%$). The above result agree with the theoretical expectations
\cite{wusmartin}. Since the mass of the quark sets a limit on the
size of the $c\overline{c}$ dipole, it becomes color transparent
and one expects a strong suppression for this configuration. On
the other hand, the effective gluon dipole, associated with
$c\overline{c}g$ production is not restricted in size.

Concerning the Regge inspired approaches, diffraction dissociation
of virtual photons furnishes the details on the nature of the
Pomeron and  on its partonic structure. As a first investigation, we follows the
Capella-Kaidalov-Merino-Tran Thanh Van (CKMT)  model to
diffractive DIS based on Regge theory \cite{CKMT1,CKMT2} and the
Ingelman-Schlein ansatz, which  is based on the intuitive picture
of a Pomeron flux associated with the proton beam and on the
conventional partonic description of the Pomeron-photon collision.
In this case, deep inelastic diffractive scattering proceeds in
two steps (the Regge factorization): first a Pomeron is emitted
from the proton and then the virtual photon is absorbed by a
constituent of the Pomeron, in  the same way as the partonic
structure of the hadrons.  In the CKMT model the structure
function of the Pomeron, $F_{\pom}(\beta,Q^2)$, is associated to
the deuteron structure function. The  Pomeron is considered as a
Regge pole with a trajectory $\alpha_{\pom}(t)$ determined from
soft processes, in which absorptive corrections (Regge cuts) are
taken into account.  The diffractive contribution to DIS is
written in the factorized form,
\begin{eqnarray}
F_2^{D(4)}(x_{\pom},\beta, Q^2,t)=
f(x_{\pom},t)\,F_{\pom}\,(\beta,\,Q^2)\,\,\,,
\label{regfac}\end{eqnarray}
 where the first factor represents the
pomeron flux from the proton and can be written as
\begin{eqnarray}
 f(x_{\pom},t) = \frac{[g^{\pom}_{pp}(t)]^2}{16
\pi}x_{\pom}^{1-2\alpha_{\pom}(t)}\;, \label{f2dregge}
\end{eqnarray}
where  $g^{\pom}_{pp}(t)=g^{\pom}_{pp}(0)\,\exp (C\,t)$ is the
Pomeron-proton coupling, with $[g^{\pom}_{pp}(0)]^2=23$ mb and
$C=2.2$ GeV$^{-2}$ \cite{CKMT1,CKMT2}. The Regge factorization
implies that the $x_{\pom}$ dependence is completely separated
from  the $\beta$ dependence, with the behavior in $x_{\pom}$
determined only by the flux factor. The value of
$\alpha_{\pom}(t)$ in the flux is given by
\begin{eqnarray}
\alpha_{\pom}(t) = 1 + \Delta (Q^2_{eff}) + \alpha^{\prime}\,
t\,\,,
\end{eqnarray}
where $\alpha^{\prime} = 0.25$ GeV$^{-2}$ and
\begin{eqnarray}
\Delta (Q^2) = \Delta (0) \left(1 + \frac{d_0 \, Q^2}{Q^2 + d_1}
\right) \,\,,
\end{eqnarray}
with $\Delta (0) = 0.09663$, $d_0 = 1.9533$ and $d_1 = 1.1606$
\cite{kaidalov_slope}. The $Q^2$ dependence of the effective
Pomeron intercept is one of the main feature of the CKMT model. It
was argued in the Refs. \cite{CKMT1,CKMT2} that this is due to the
fact that the size of the absorptive corrections decreases when
$Q^2$ increases. This parameterization gives a good description of
all existing data on $\gamma^* p$ total cross section in the
region $Q^2 \le 10$ GeV$^2$ \cite{kaidalov_slope}. At larger
$Q^2$, effects due to QCD evolution become important.  The scale
$Q^2_{eff}$ is a priori not known. From a theoretical point of view,
values for $\Delta (Q^2_{eff})$ between 0.13 and 0.24 are
possible, corresponding to the effective Pomeron intercept without
eikonal-type corrections and the "bare" value, respectively. Both
values are not excluded by the recent fit for the data which
assumes in addition to the Pomeron exchange, the contribution of a
subleading reggeon trajectory \cite{royon1,royon2}. Integrating
Eq. (\ref{regfac}) over $t$, $F_2^{D(3)}$ can be put in the
factorized form
\begin{eqnarray}
F_2^{D(3)}(x_{\pom},\beta, Q^2)=
\overline{f}(x_{\pom})\,F_{\pom}\,(\beta,\,Q^2)\,\,\,,
\label{regfac2}
\end{eqnarray}
where $\overline{f}(x_{\pom})$ is the $t$-integrated pomeron flux
\begin{eqnarray}
\overline{f}(x_{\pom}) = \int_0^{\infty} d|t| \,f(x_{\pom},t)\,\,.
\label{fluxint}
\end{eqnarray}
It must be stressed that since the Pomeron is not a particle the
separation of the flux factor from the photon-Pomeron cross section is quite
arbitrary, and therefore the  normalization of the flux is ambiguous.

The second factor in Eq. (\ref{regfac}) is the pomeron structure
function $F_{\pom}$ and is proportional to the virtual
photon-pomeron cross cross section.
  In
the CKMT approach, $F_{\pom}(\beta,Q^2)$ is determined using Regge
factorization and the values of the triple Regge couplings
determined from soft diffraction data. Namely, the Pomeron
structure function is obtained from $F_2^p$, or more precisely
from the combination $F_2^d=\frac{1}{2}(F_2^p + F_2^n)$, by
replacing the Reggeon-proton couplings by the corresponding triple
reggeon couplings (see Ref. \cite{CKMT1} for details). The
following parametrization of the deuteron  structure function
$F_2^d$ at moderate values of $Q^2$ (and small-$x$), based on
Regge theory, was introduced,
\begin{eqnarray}
 F_2^d(x, Q^2)  =   A  \,
x^{- \Delta(Q^2)}(1 - x)^{n(Q^2)+4} \left ( {Q^2 \over Q^2 + a} \right )^{1 +
\Delta(Q^2)} ,
 \end{eqnarray}
where $1 + \Delta(Q^2)$ is the  Pomeron intercept, which depends
on the photon virtuality. The Pomeron structure function is
identical to $F_2^d$ except for a simple changes in its
parameters: $F_{\pom}(\beta, Q^2) = F_2^d \left (x \to  \beta; A
\to eA, n(Q^2) \to n(Q^2) - 2 \right )$.  The value of  $e$ in
$F_{\pom}$ is obtained from conventional triple reggeon fits to
high mass single diffraction dissociation for soft hadronic
processes. The remaining  parameters  are given in Refs.
\cite{CKMT1,CKMT2}. In the CKMT approach, the gluon distribution
of the Pomeron can be obtained for low $\beta$ in a similar way as
for the quarks discussed above. It is written as,
\begin{eqnarray}
\beta\,g_{\pom}\,(\beta,Q^2) = e_d^{\pom}\,C_g\,\beta^{-\Delta (Q^2)}\,(1-\beta)^{n_g}\,,
\label{gluondregge}
\end{eqnarray}
where $n_g$ is a free parameter  and $e_d^{\pom} = r_{\pom\,
\pom}^{\pom}(t)/g_{dd}^{\pom} =0.07$, with $r_{\pom \pom}^{\pom}$
and $g_{dd}^{\pom}$ being the couplings of the Pomeron to the
Pomeron and to the deuteron, respectively. The distribution is
singular towards $\beta \to 0$ due to the powerlike behavior
driven by the Pomeron exchange. In Ref. \cite{haakman}, where
charm diffractive production was computed, $n_g$ takes values
between 0 and -1  in order to produce a normalizable distribution,
which implies for $n_g < 0$ a singular behavior also at $\beta=1$.
As a good description of the $Q^2$ dependence of the HERA data is
achieved with  $n_g=0$ and in view that a singular behavior for
large $\beta$ is not observed in Regge-like fitting procedures to DDIS
data,  we assume this value in our analyzes.

In the Regge based approaches,  the massive charm contribution
arises from photon-gluon fusion. The diffractive structure
function is  $F_2^{D(3)\,\mathrm{charm}} =
\overline{f}(x_{\pom})\times F_{\pom}^{c\bar{c}}(\beta,Q^2)$,
where $\overline{f}(x_{\pom})$ is given by Eq. (\ref{fluxint}) and
the charmed contribution to the diffractive Pomeron structure
function, $F_{\pom}^{c\bar{c}}(\beta,Q^2)$,  is given by folding
the gluon distribution Eq. (\ref{gluondregge}) in Eq.
(\ref{f2dqqbarg}) \cite{haakman}. The factorization scale is
assumed equal to $4\,m_c^2$. In the general case, the scaling
violations of Pomeron structure function  should be considered.
However, as the $Q^2$-range of the HERA data considered here is either
 small and the
parameters in (\ref{gluondregge}) have been obtained for $Q^2 = 5$
GeV$^2$, in a first approximation we disregard the logarithmic
dependence on $Q^2$, which is given by QCD-evolution.

Another possible approach is the  QCD analysis of the diffractive structure function in terms of both Regge phenomenology and perturbative QCD evolution as made in Ref. \cite{royon1}. In this case the parton distributions of the Pomeron are derived  from QCD fits of diffrative deep inelastic scattering cross sections determined at HERA. In particular, the diffractive structure functions is given by
\begin{eqnarray}
F_2^{D(3)}(Q^ 2,\beta, x_{\pom}) = f_{\pom/p}(x_{\pom})F_2^{\pom}(Q^2,\beta) + f_{\reg/p}(x_{\pom})F_2^{\reg}(Q^2,\beta)  \,\,,
\end{eqnarray}
where $F_2^{\pom}$ can be interpreted as the Pomeron structure function and $F_2^{\reg}$ as an effective Reggeon structure function, with the restriction that it takes into account various secondary Regge contributions which can  hardly  be separated. The  Pomeron and Reggeon fluxes are assumed to follow a Regge behavior with linear trajectories $\alpha_{\pom,\reg}(t)=\alpha_{\pom,\reg}(0)+\alpha^{\prime}_{\pom,\reg}t$, such that
\begin{eqnarray}
 f_{\pom/p, \reg/p}(x_{\pom}) = \int_{t_{cut}}^{t_{min}} \frac{e^{B_{\pom,\reg}t}}{x_{\pom}^{2 \alpha_{\pom,\reg}(t) - 1}} dt \,\,,
\label{fluxbar}
\end{eqnarray}
where $|t_{min}|$ is the minimum kinematically allowed value of $|t|$ and $t_{cut}=-1$ GeV$^2$ is the limit of the measurement.  The gluon distribution for the Pomeron is parameterized in terms of non-perturbative input distributions at $Q_0^2= 3$ GeV$^2$ as follows
\begin{eqnarray}
\beta G(\beta, Q^2 = Q_0^2)=\left[\sum_{j=1}^n C_j^{(G)}P_j(2z-1)\right]^2 \,e^{\frac{a}{\beta-1}}\,\,,
\end{eqnarray}
and similarly for the quark flavor singlet distribution. The $P_j(\eta)$ is the $j^{th}$ member in a set of Chebyshev polynomials, which are chosen such that $P_1 = 1$, $P_2 = \eta$ and $P_{j+1}=2\eta P_j(\eta) - P_{j-1}$.  Here we consider this parameterization for the gluon distribution  and the corresponding Pomeron flux [Eq. (\ref{fluxbar})] as input in our calculations. The parameters used are from the H1 fit in Ref.\cite{royon1}.

\section{Results and discussion}

\begin{figure}[t]
\centerline{\epsfig{file=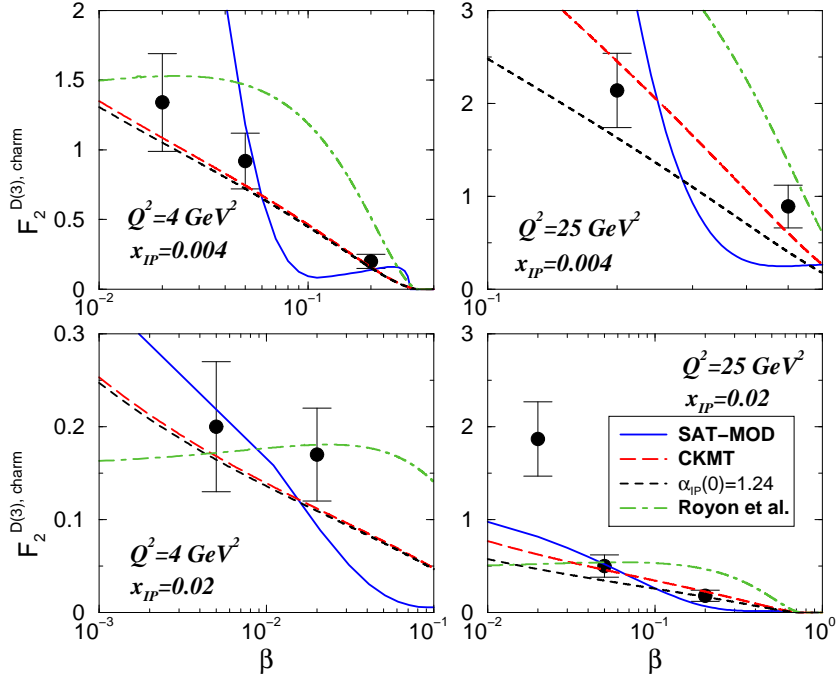,width=110mm}} \caption{The
open charm diffractive structure function
$F_2^{D(3)\,\mathrm{charm}}$ as a function of $\beta$ (data from
ZEUS Collaboration \cite{zeus}). The solid lines correspond to the
perturbative QCD calculation, whereas the other lines represents
the results from the Regge approach. The long-dashed curves stand for
the CKMT Pomeron and the dashed ones for a fixed Pomeron intercept
(see text). The dot-dashed curves represents the result using a
Regge/QCD analysis from Ref. \cite{royon1}}
 \label{fig:2}
\end{figure}

In the previous section,  we have  reviewed the formulas for the
open-charm contribution to the proton diffractive structure
functions in the perturbative QCD formalism and Regge based
approach.  In that follows, one  computes the charm diffractive
structure function considering these  different analysis without additional parameters. In Fig.
\ref{fig:2} one presents the results  for the QCD approach (solid
lines), the CKMT model (dashed and long-dashed lines) and the QCD analysis from Royon et al. \cite{royon1} (dot-dashed lines) using $m_c=1.5$ GeV. In
particular, we consider two possibilities for the effective
Pomeron intercept $\alpha_{\pom}(0) = 1 + \Delta (Q^2_{eff})$, 
which determines the $x_{\pom}$ dependence of the Pomeron flux.
Basically, we have considered the  higher ($\alpha_{\pom}(0)=1.24$)
bound obtained in the HERA fit and  also an intercept  $Q^2$-dependent
(the CKMT Pomeron).

Regarding the CKMT approach, as the parameters have been constrained for $Q^2 = 5$ GeV$^2$, we initially compare our predictions with the experimental results for $Q^2 = 4$ GeV$^2$. We have that Regge based approach agrees with ZEUS data both
in shape and overall normalization for $Q^2$ dependent Pomeron
intercept  and/or fixed $\alpha_{\pom}=1.24$. For larger $Q^2$ we have
that these two choices give  different normalizations. However, due to
the scarce data a discrimination is still not possible. Moreover, this
result can be modified by the QCD evolution, which is not considered
in our analyzes.   On the other hand, the $\beta$ dependence predicted by the CKMT approach is consistent with the behavior present in the experimental measurements.  
This result is supported by  the
phenomenological analyzes from ZEUS \cite{zeus}, which uses a fitting
procedure based on QCD factorization for diffractive DIS in order to
determine the diffractive quark and gluon distributions.

The result when  using the gluon distribution from Ref. \cite{royon1}
is quite different from that one coming from the  CKMT model. In
particular, the deviation is increasingly larger at  small $x_{\pom}$
and large $Q^2$. Moreover, the behavior at small $\beta$ has changed,
which becomes flat at this kinematical region. The reason for that is
an almost flat diffractive gluon distribution at small $\beta$ coming out of the fit of
Ref. \cite{royon1}, whereas one has a singular behavior when
considering the CKMT gluon distribution on the Pomeron. These results strongly
indicate that the charm diffractive production could allows us further
 constrain future analysis on the diffractive parton (gluon) 
distributions  on diffractive DIS. It should be noticed a new set of
NLO DGLAP/QCD diffractive parton distributions has been recently determined
\cite{H1diffnew} (preliminary), which includes for the first time both
experimental and model uncertainties for the error bands of the
diffractive pdf's.  There, the small $\beta$ behavior is steep in
contrast with the almost flat gluon distribution found in the fit of
Ref. \cite{royon1}. Therefore, it is expected the results using these
new parton distributions will modify the analysis presented here.

The perturbative QCD approach  provides a steep behavior on
$\beta$ in comparison with the Regge based one. 
In particular, for small $\beta$ and small $x_{\pom}$ the difference between the predictions is sizeable.
The main
contribution in the pQCD approach comes from the $c\bar{c}g$ component, which is
strongly dependent on the input for the diffractive gluon
distribution. Moreover, the implicit dependence on $\beta$ 
present in the upper limit of the $k_t$-integration in  Eq. (\ref{GD}) implies an additional  $\beta$ dependence. For instance, if we assume in a first approximation that $x_{\pom}g_{\pom}(x_{\pom} , k_t^2) \propto \ln k_t^2$, it would produce a logarithmic enhancement in this dependence.    
Accordingly, the numerical calculation of Eq. (\ref{GD}) produces a
strong growth at small $\beta$ for $x_{\pom}=0.004$ at both virtualities, either
overestimating the data points. However, the description is in
agreement with data for   $x_{\pom}=0.02$, even at high $Q^2$. 
It should be noticed that the QCD evolution in the unintegrated gluon distribution may modify this scenario. Furthermore, it is important to emphasize that the pQCD approach predicts a quadratic dependence on $x_{\pom}g_{\pom}$. Consequently, for a typical powerlike behavior we expect a stronger dependence  than 
  present in the Regge models. Therefore, a better  discrimination between the models can be obtained by increasing the data statistics and enlarging the kinematical window.

As a summary, it was shown the diffractive production of open charm is
an important  observable testing QCD dynamics. The ZEUS  collaboration has recently measured \cite{zeus} the open
charm diffractive   structure function
$F_2^{D(3)\,\mathrm{charm}}$, which is extracted from charmed
mesons $D^{*\pm}(2010)$ production. The data demonstrate a strong
sensitivity to the diffractive parton densities. 
 Here, we have contrasted the pQCD two-gluon exchange
approach and  Regge/QCD models.  For the first one, the saturation model
was considered in order to write down the unintegrated gluon
distribution. In this case was observed good description at larger 
$x_{\pom}$, but a sizeable underestimation at smaller values of
$x_{\pom}$, mostly at smaller $\beta$, is verified. Concerning 
Regge approach, the  CKMT model gives a reasonable data description in
shape and normalization with/without a $Q^2$-dependent Pomeron
intercept. On the other hand, the Regge/QCD approach of
Ref. \cite{royon1} provides a flat behavior at small $\beta$, which is
not consistent with the current measurements. A comparison with the most
recent parameterizations of diffractive pdf's is timely. 
We  conclude that an increasingly
experimental statistics on this process would help to constrain the
diffractive gluon distribution appearing in diffractive factorization
approaches and/or discriminate among several parameterizations for the
gluon distribution in the Pomeron in approaches based on Regge
phenomenology.

\section*{Acknowledgments}
M.V.T.M. is grateful for the  hospitality and financial support of
CERN Theoretical Physics Division,  where part of this work was
performed.  The authors would like to thank Prof. Alan Martin
(Durham U., IPPP and Canterbury U.) for helpful comments. This
work was partially supported by CNPq, Brazil.

\end{document}